\documentclass[11pt]{article}
\usepackage{hyperref}
\pdfoutput=1

\begin{document}

\title{Dynamic  States in Cellular Flames}

\author{Michael Gorman \\
\\vspace{6pt} Department of Physics\\University of Houston\\
Houston, TX 77204-5005}

\maketitle


\begin{abstract}
A premixed flame on a circular porous plug burner can form cellular flames, 
which are ordered patterns of concentric rings of brighter, hotter regions
separated by darker, cooler cusps and folds.  These ordered states 
bifurcate to dynamic states in which the cells move in coordinated motion.  Fluid
dynamics videos of representative examples of these states will be presented.

\end{abstract}




\href{http://hdl.handle.net/1813/13745}{Dynamic States of Cellular Flames.mpg}

A premixed flame on a circular porous plug burner appears a thin ($0.5$ mm) luminous 
disk.  Premixed flames are distinctive as a dynamical system in that both the spatial and 
temporal characteristics of the can be measured to high resolution.   Cellular flames of 
rich hydrocarbon-air mixtures form ordered states of concentric rings of brighter, hotter 
cells separated by darker, cooler cusps and folds.  At larger values of the control 
parameters, the ordered states are observed to bifurcate to dynamic states: 1) rotating states 
in which rings of cells undergo rotation ($\sim 100$ deg/sec), 2) hopping states in which individual 
cells abruptly change their angular position in a hopping-like motion, 3) intermittently 
ordered states in which ordered patterns intermittently appear and disappear at irregular 
intervals, and 4) ratcheting states in which rings of cells rotate very slowly ($\sim 1$ deg/sec), 
speeding up and slowing down in a manner characteristic of each state.  Fluid dynamics videos 
of each state will be shown.  This dynamics will be discussed in the language of dynamical systems.


%
\end{document}